# Cascade spin dynamics of excitons localized in indirect-band-gap (In,Al)As/AlAs quantum dots with type-I band alignment

**S. V. Nekrasov** [1,*], I. V. Kalitukha[1], A. A. Golovatenko[1], Ya. A. Kuznetsova[1], N. O. Mikhailenko[1], M. D. Ragoza[1], T. S. Shamirzaev[2] and Yu. G. Kusrayev[1]

[1] *Ioffe Institute, Russian Academy of Sciences, 194021 St. Petersburg, Russia*

[2] *Rzhanov Institute of Semiconductor Physics, Siberian Branch of the Russian Academy of Sciences, 630090 Novosibirsk, Russia*

**e-mail:* sergey.nekrasov@mail.ioffe.ru

We investigate the spin dynamics of excitons localized in type-I (In,Al)As/AlAs quantum dots with an indirect in momentum space band structure. Polarized selective photoluminescence spectroscopy, i.e. fluorescence line narrowing, under magnetic fields up to 5 T applied in the Faraday geometry is employed. The experiment reveals a cascade spin evolution process of excitons in the indirect-band-gap quantum dots: an initial short-term spin dynamics associated with excited direct exciton states possessing a large oscillator strength is followed by electron relaxation into the X-valley of the Brillouin zone and subsequent long-term spin dynamics of indirect excitons. The two-step mechanism manifests itself in the distinct features of the magnetic field dependences of photoluminescence: two-component recovery of optical orientation, two-component linear-to-circular polarization conversion and the presence of the linear polarization plane rotation. At the same time, suppression of the optical alignment shows one-component behavior governed by the spin dynamics of the indirect exciton states. Within the pseudospin formalism, we derive analytical expressions that quantitatively describe the observed dependences and yield estimates for the anisotropic exchange splitting: 210 μeV for direct excitons and 1.3 μeV for indirect excitons. Further analysis using the density matrix formalism agrees well with the pseudospin model calculations and shows that the finite optical orientation at zero magnetic field is due to comparable magnitudes of the anisotropic splitting of the indirect exciton states and the splitting of the X-valley electron states caused by the hyperfine interaction with nuclei.

## I. INTRODUCTION

The spin of an electron confined in a semiconductor quantum dot (QD) is a fundamental model object in spintronics [1-4]. In QDs, the main mechanisms of carrier spin relaxation that are relevant in bulk materials, in particular the Dyakonov–Perel mechanism, are suppressed [1, 5]. At the same time, the exchange interaction in the exciton localized in a QD is enhanced compared with that in a bulk material [1]. The anisotropic component of the exchange interaction leads to a rapid loss of exciton spin orientation on a timescale of about 10 ps for typical values of the anisotropic exchange splitting of about 100 μeV [6-8]. In addition, the lifetime of the photoexciton in direct-band-gap QDs, which is typically on the order of 1 ns [9], is too short for practical use of spin in information processing and storage devices. In this context, QDs with the indirect band gap in momentum space are promising, since the exciton lifetime in such structures can reach milliseconds [10], while the spin relaxation time may even exceed the lifetime [11-13]. Indirect-band-gap QDs are an interesting model system because they can exhibit new spin phenomena. In particular, in (In,Al)As/AlAs QDs, dynamic polarization of electron spins interacting with nuclei [14], blocking of light absorption due to the population of the QDs by dark excitons [15] and anisotropy of the hyperfine interaction of electrons and nuclei [16] were observed.

It was previously shown that, depending on their size, (In,Al)As/AlAs QDs with type-I band alignment can be either direct or indirect in momentum space [11, 17]. In relatively large QDs, the lowest energy state in the conduction band is located in the $\Gamma$-valley of the Brillouin zone, while in small QDs it is located in the X-valley. This feature arises from the fact that the electron effective mass in the X-valley is significantly larger than that in the $\Gamma$-valley [17]. In agreement with observations in various direct-band-gap QD systems [6-8], the exciton spin dynamics in direct-band-gap (In,Al)As/AlAs QDs is governed by the exchange interaction of the electron and the hole, with the anisotropic exchange splitting of the exciton states $\delta_1$ of several hundreds of μeV [11, 18]. In QDs with the band gap indirect in momentum space, the exchange interaction is weakened to values of $\delta_1 \leq 1$ μeV [11, 19], because of the weak overlap in the momentum space between the wave functions of the $\Gamma$-valley hole and the X-valley electron. As a result, the exciton spin dynamics can be governed by both the exchange interaction [19] and the hyperfine interaction of the X-valley electron with nuclei [11, 14, 20, 21]. The small exchange splitting of the exciton states makes indirect-band-gap (In,Al)As/AlAs QDs a promising system for the generation of entangled photons via the biexciton–exciton radiative cascade [22, 23].

In this work, a subensemble of indirect-band-gap (In,Al)As/AlAs QDs with type-I band alignment was studied using selective optical excitation. It was previously established [19] that this subensemble is characterized by significant $\Gamma$-X mixing of the electron states. As shown in Fig. 1(a), in such QDs the light predominantly excites the direct $\Gamma$-exciton, which has a relatively large oscillator strength. This is followed by electron energy relaxation from $\Gamma$ into the X-valley of the Brillouin zone on a timescale of a few picoseconds [24], which enables efficient optical excitation of the indirect exciton states. We investigate the role of intermediate $\Gamma$-states in the spin dynamics of excitons in indirect-band-gap QDs. The optical orientation and optical alignment effects are studied subject to Faraday magnetic field over a wide range of magnetic fields up to 5 T. Photoluminescence (PL) polarization conversion effects are also measured. In Sec. IV, we propose a model based on a two-step process of exciton spin evolution that describes the unusual dependences observed experimentally.

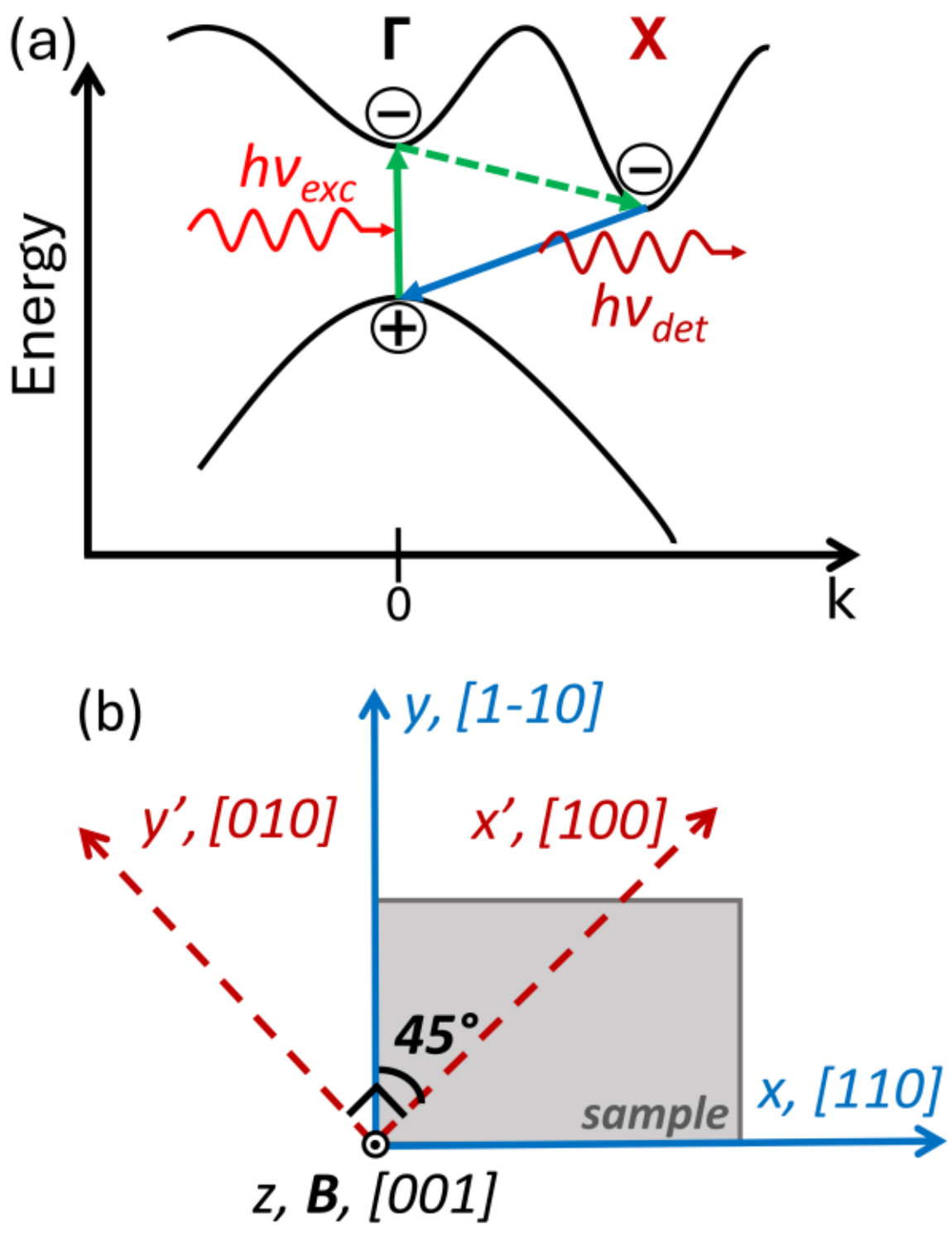

FIG. 1. (a) Diagram of an exciton life cycle in the (In,Al)As/AlAs QDs. Absorption of the photon leads to the excitation of the direct states (green upward arrow), followed by the electron relaxation to the X-valley of the Brillouin zone (green dashed arrow) and radiative recombination (blue arrow). (b) Geometry of the optical experiment with the magnetic field ***B*** applied along the z axis. Here z is co-directed with the structure growth axis [001], x axis is co-directed with the [110] crystallographic direction, y - with [1-10], the x' and y' axes are inclined at an angle of $45^{\circ}$ relative to the x and y axes.

## II. EXPERIMENTAL DETAILS

### A. Sample

The sample grown by the molecular beam epitaxy contains 20 layers of (In,Al)As QDs. The average size of the QDs is 15*15*4 $nm^3$ with a diameter dispersion of 3.25 nm, here we address relatively large QDs by means of selective optical excitation. The density of the QDs in a layer is about $3*10^{10}$ $cm^{-2}$, the layers are separated by AlAs barriers with a thickness of 20 nm which prevent the interaction of quasi-particles in adjacent layers of QDs [25, 26]. For more details about the growth process see Ref. [17].

### B. Experimental setup

The ensemble of (In,Al)As/AlAs QDs was studied using optical orientation and optical alignment techniques in Faraday magnetic field. Different types of PL polarization conversions were measured as well. The excitation axis was chosen along the growth direction of the structure [001], as shown in Fig. 1(b). The sample was mounted in the variable temperature insert of a helium-bath cryostat equipped with a split-coil superconducting solenoid and immersed in pumped liquid helium at a temperature of 1.6 K. The magnetic field of up to 5 T was applied in the Faraday geometry along the z axis, see Fig. 1(b). PL was excited quasi-resonantly using a Ti:sapphire laser with a photon energy of 1.66 eV. The PL collected in the backscattering geometry was dispersed by the U-1000 double monochromator (Jobin Yvon) with 500 μm slits providing overall spectral resolution of 0.2 meV. The PL was detected by a GaAs photomultiplier operating in photon-counting mode.

We used the following notations to describe experimental conditions: indexes *a* and *b* of the PL polarization degree ($P_a^b$) denote the type (linear or circular) of the polarization of the detected luminescent emission and the exciting laser light, respectively. The circular polarization of the PL is defined as:

$$P_c^b = \frac{I^+ - I^-}{I^+ + I^-}, \quad (1)$$

where $I^+$ ($I^-$) is the intensity of the PL component with $\sigma^+(\sigma^-)$ polarization. The following experiments were conducted: the degree of PL circular polarization was measured under excitation with circularly polarized light ($P_c^c$), i.e. the optical orientation effect; the degree of PL circular polarization under excitation with the light linear polarized along the x axis ($P_c^l$, see Fig. 1(b)), i.e. the conversion from linear polarization to circular.

The linear polarization of the PL is defined as:

$$P_l^b = \frac{I^{[110]} - I^{[1-10]}}{I^{[110]} + I^{[1-10]}}, \quad (2)$$

where $I^{[110]}$ ($I^{[1-10]}$) is the intensity of the PL component linearly polarized along the [110]( [1-10]) direction. The linear polarization of the PL was measured under linearly polarized excitation along the [110] direction ($P_l^l$, optical alignment) or along the [100] direction ($P_l^{l'}$, rotation of the polarization plane).

In Model section we introduce PL polarization $P_a^{b_[ijk]}$ calculated for QDs elongated along the specific crystallographic direction [*ijk*], directions [110], [1-10], [100] and [010] are considered.

## III. EXPERIMENTAL RESULTS

Fig. 2 shows the PL spectrum of (In,Al)As/AlAs QDs under selective excitation. By means of the fluorescence line narrowing [27, 28], a subensemble of indirect-band-gap QDs with a spectral width of 8 meV was addressed. The feature at 1.646 eV in the spectrum is likely due to recombination assisted by a transverse acoustic AlAs phonon with the characteristic energy of 13 meV [29]. The spectral width of the entire inhomogeneously broadened QD ensemble is about 200 meV [11], see Fig. A(b) in Appendix. Fig. 2 also presents the spectral dependences of the optical orientation and optical alignment effects. At the maximum of the PL band, the optical orientation increases from 11% in zero magnetic field to 81% in the magnetic field of 5 T. The optical alignment is suppressed by the magnetic field of 4 T, decreasing from 43% in zero field to 3%.

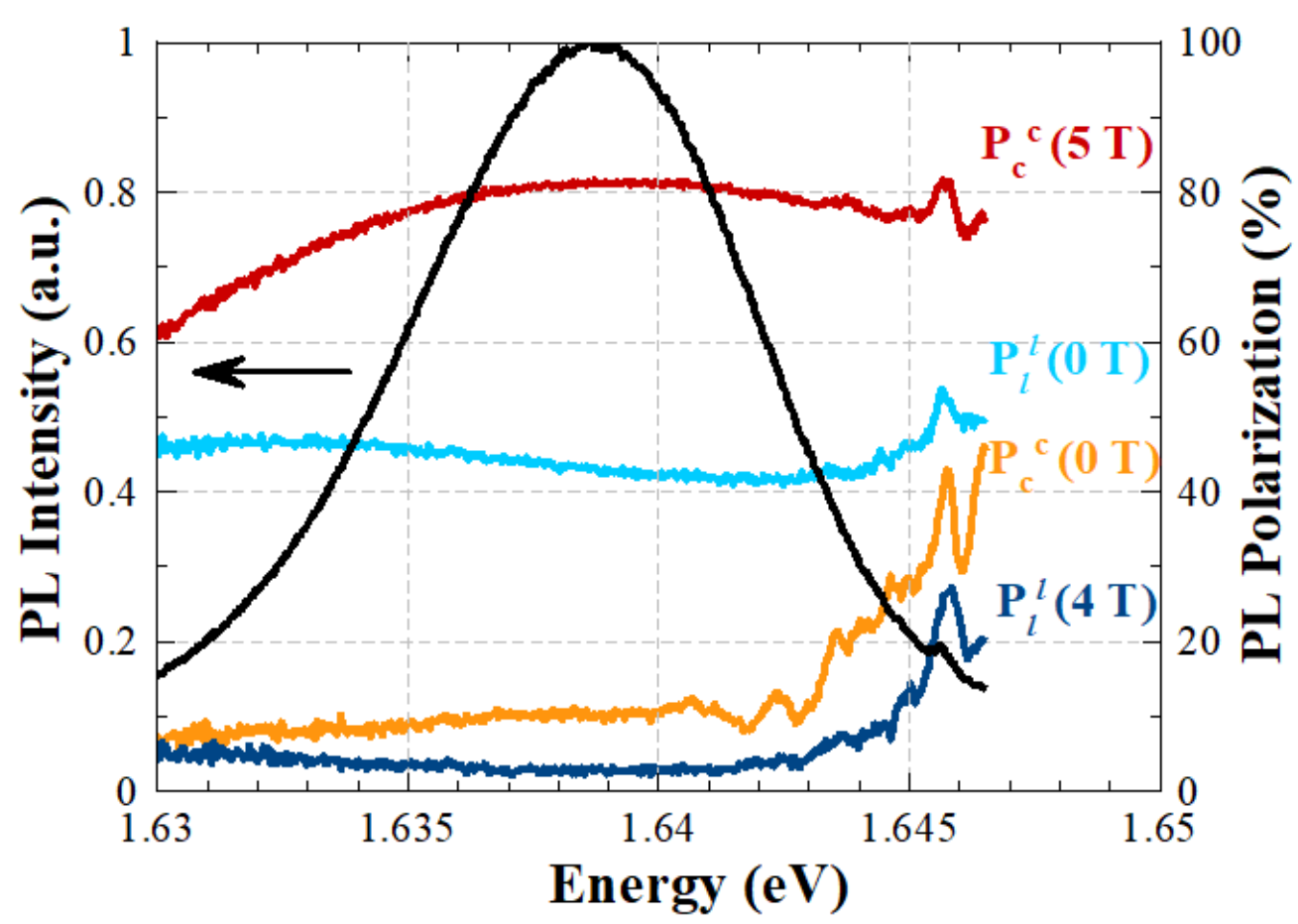

FIG. 2. PL spectrum and polarization under excitation with a photon energy of 1.66 eV. The PL spectrum has a bell-shaped profile and is shown by the black line. The luminescence linear polarization under linearly polarized optical excitation along [110] ($P_l^l$) is shown for magnetic fields of 0 and 4 T in the Faraday geometry by the cyan and blue lines, respectively. The circular polarization under circularly polarized excitation ($P_c^c$) was measured at magnetic fields of 0 and 5 T and is shown by the orange and red lines, respectively. The temperature is 1.6 K.

Fig. 3(a) shows the magnetic-field dependence of the optical orientation effect. The optical orientation is restored in magnetic field from 11 to 81%. The dependence has a two-component shape and is described by the sum of two Lorentzian contours with half widths at half maximum (HWHM) of 50 mT and 1.6 T. Fig. 3(b) shows the optical alignment effect and its suppression from 43 to 3% by magnetic field. This dependence is described by a single Lorentzian contour with a HWHM of 50 mT.

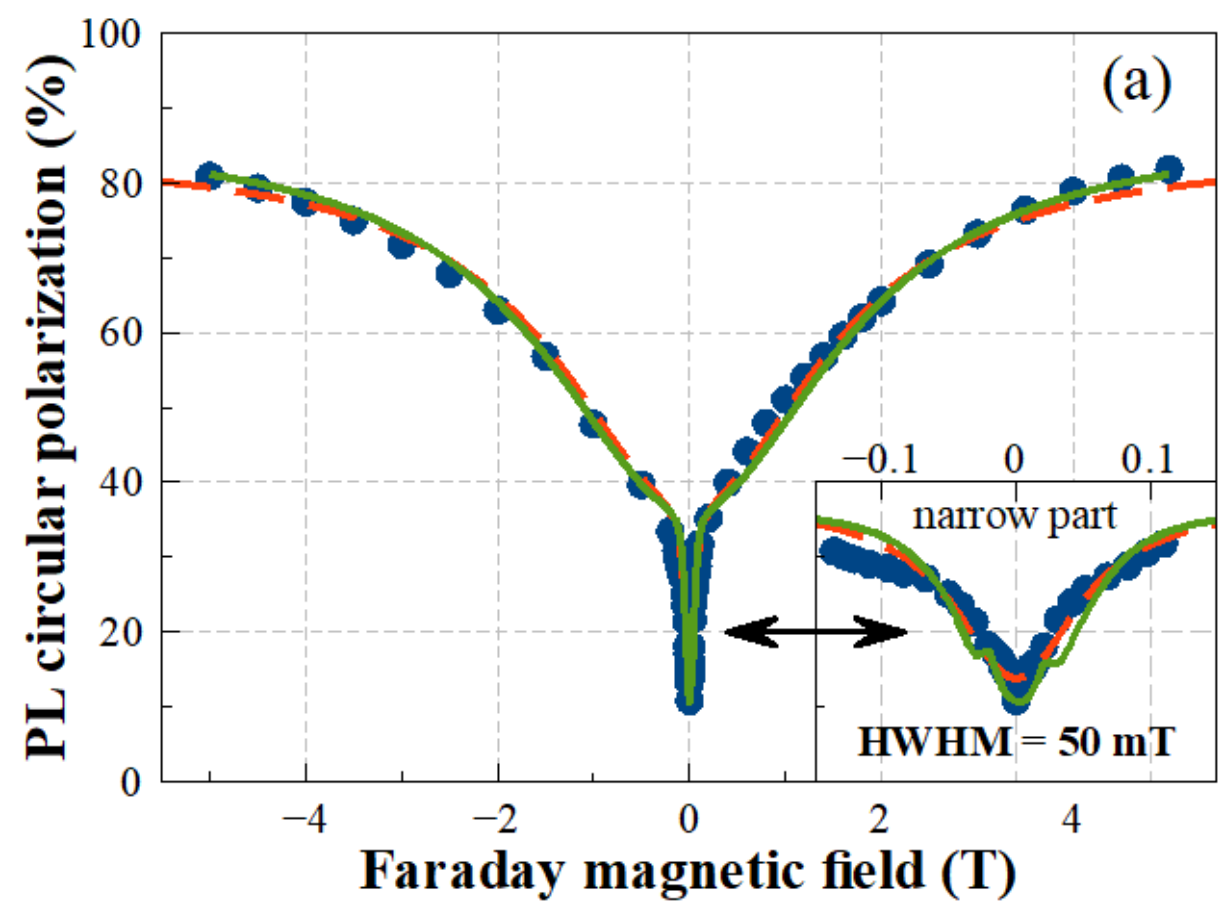

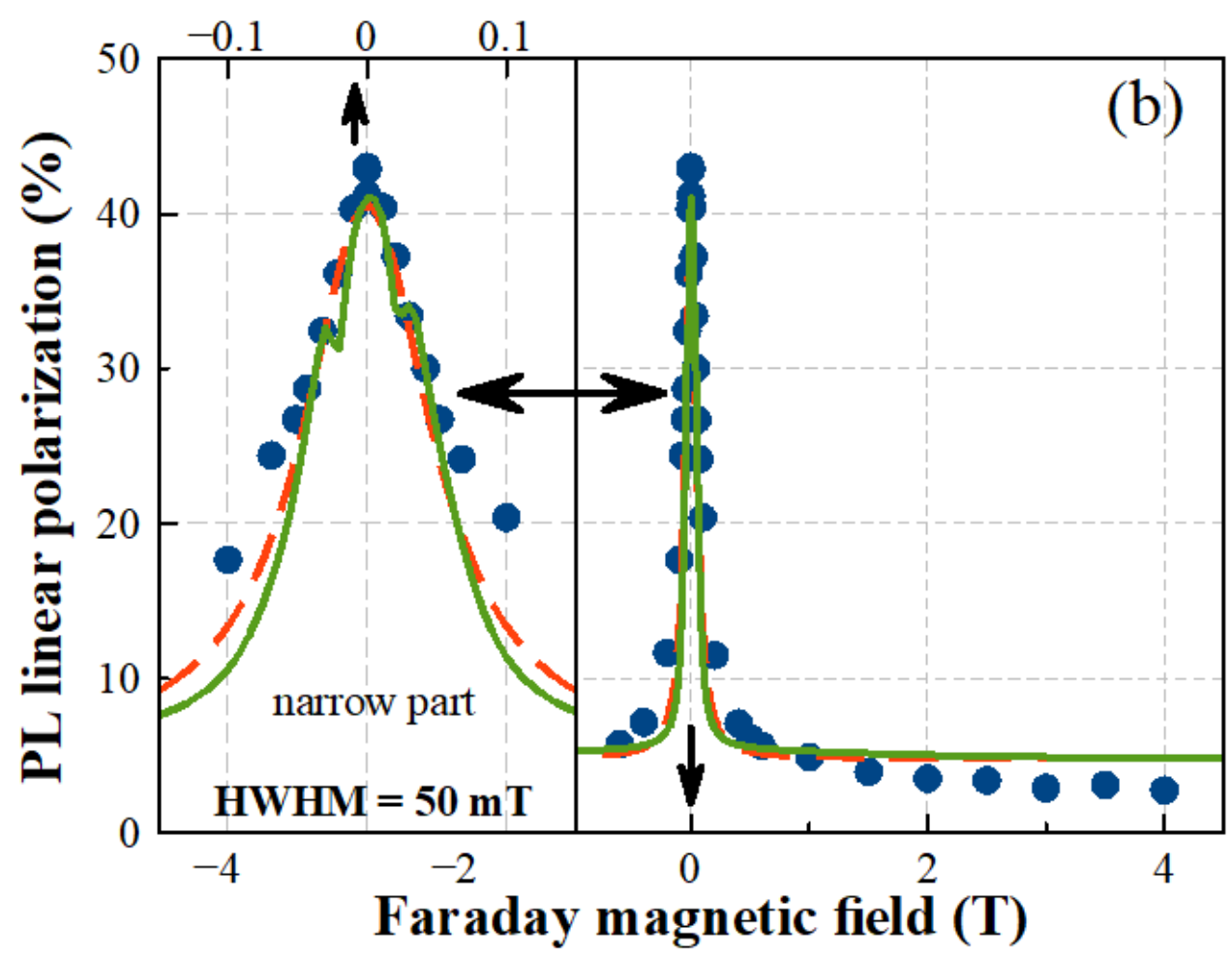

FIG. 3. (a) Circular polarization of PL under circularly polarized excitation as a function of the magnetic field in Faraday geometry. (b) Linear polarization of PL for excitation linearly polarized along [110] ( $P_l^l$ ) as a function of magnetic field in the Faraday geometry. The symbols represent experimental data, the model curves within the pseudospin formalism and density matrix formalism are shown by the dashed red lines and solid green lines, respectively. The insets show the narrow parts of the dependences in small fields from -0.15 to +0.15 T. The PL excitation energy is 1.66 eV, the detection energy is 1.64 eV. The temperature is 1.6 K.

Fig. 4(a) shows the conversion of laser light linearly polarized along the [110] crystallographic direction into circularly polarized PL with a magnitude of about 5%. At relatively low magnetic fields of about 75 mT, the conversion of linear polarization from [100] to [110] with an amplitude of about 3.5% is observed, see Fig. 4(b).

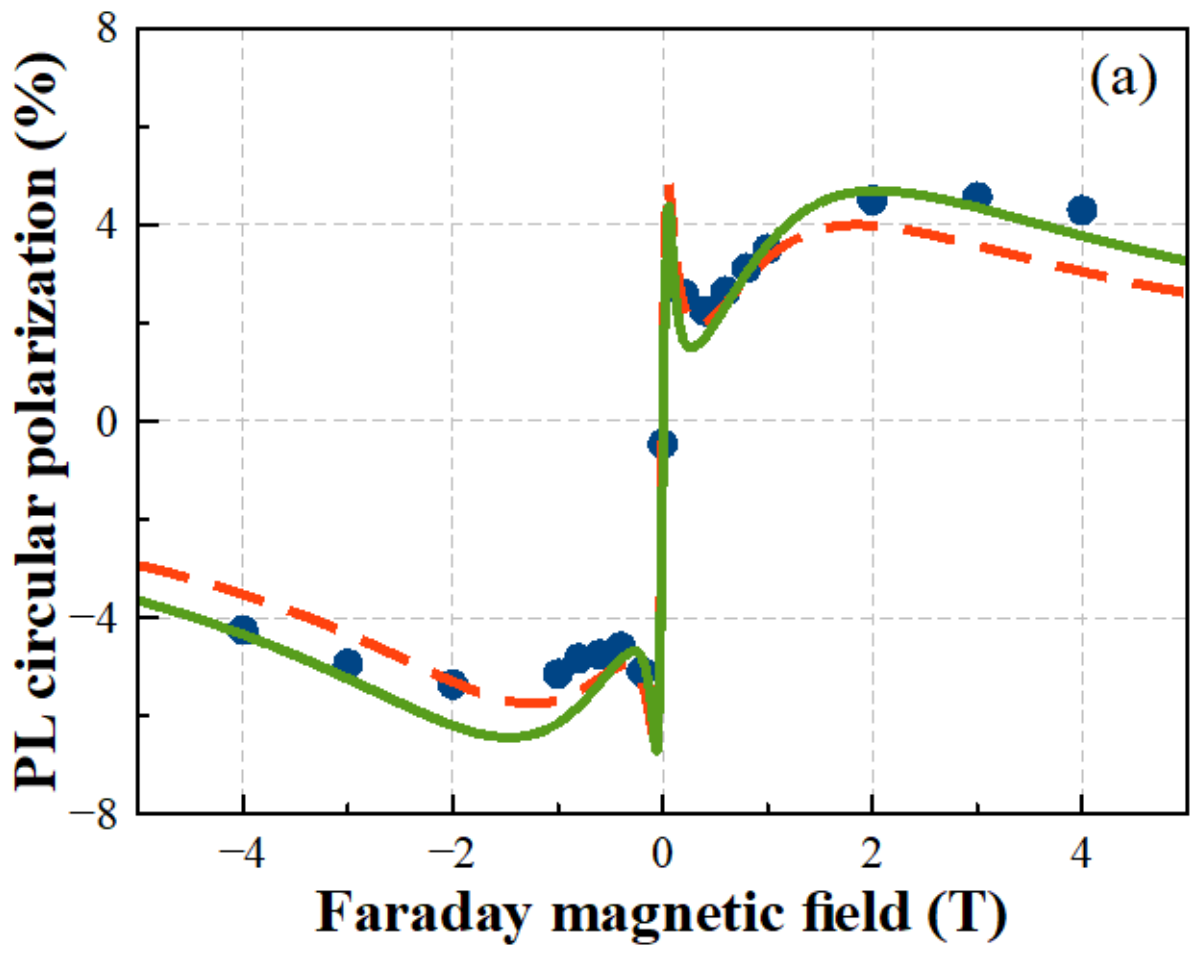

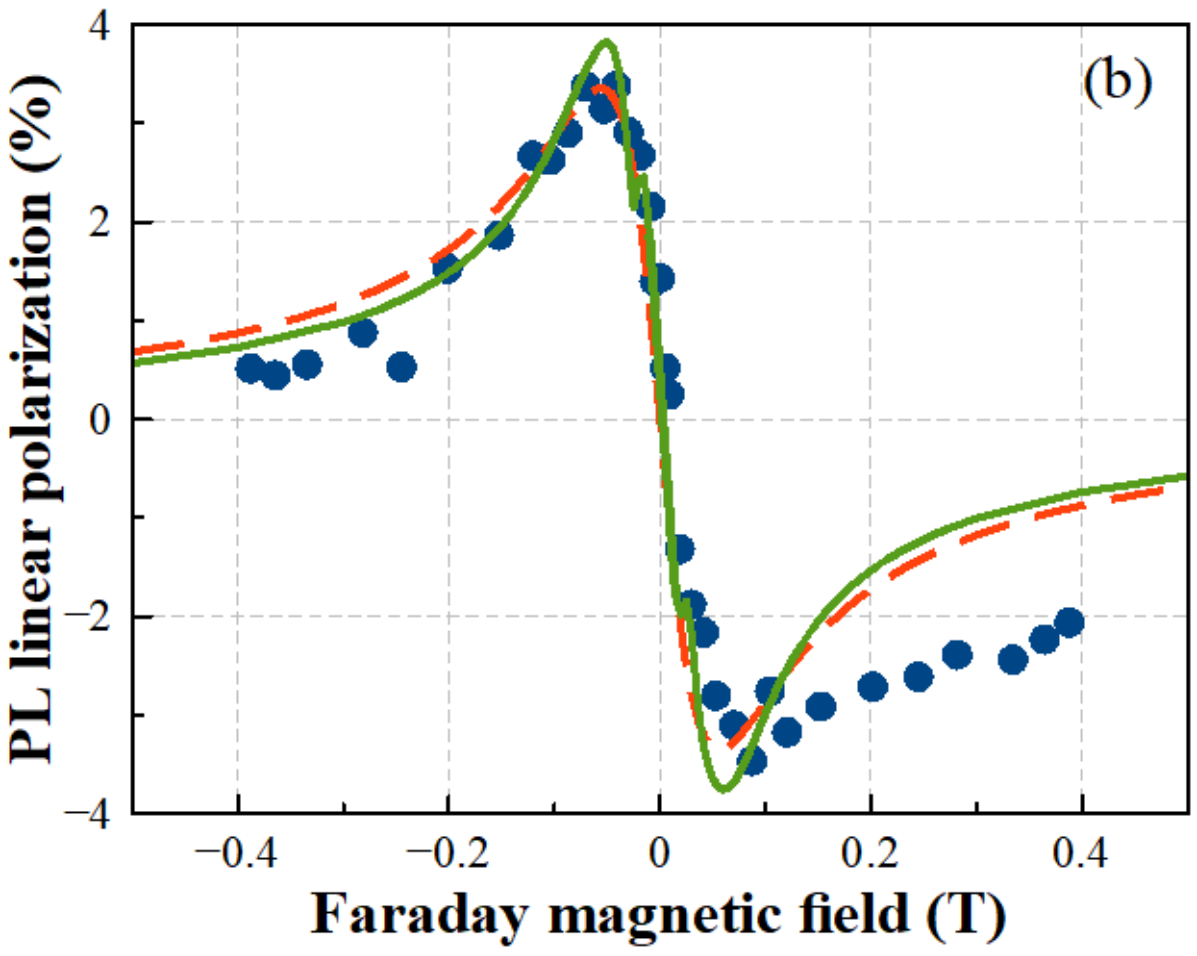

FIG. 4. (a) Circular polarization of PL under linearly polarized along [110] excitation $P_l^c$ as a function of magnetic field in the Faraday geometry. (b) Linear polarization of PL under linearly polarized along [100] excitation $P_l^{l'}$ as a function of magnetic field in the Faraday geometry. The symbols represent experimental data, the model curves within the pseudospin formalism and density matrix formalism are shown by the dashed red lines and solid green lines, respectively. The PL excitation energy is 1.66 eV, the detection energy is 1.64 eV. The temperature is 1.6 K.

## IV. MODEL

### A. Two-step spin evolution model within the pseudospin formalism

To address the distinct features of the experiment, it is necessary to take into account the spin interactions of an electron and a heavy hole localized in a QD. The isotropic component of the exchange interaction of the electron and the hole splits the exciton states into optically active doublet $|\pm1\rangle$ and optically inactive states $|\pm2\rangle$, separated by the energy $\delta_0$ [30]. Due to the violation of axial symmetry in the QD plane, further splitting of the states occurs. In particular, the optically active doublet is split into linearly polarized states $|X\rangle = {}^{1}\!/\!_{\sqrt{2}}(|+1\rangle + |-1\rangle)$ and $|Y\rangle = {}^{1}\!/\!_{i\sqrt{2}}(|+1\rangle - |-1\rangle)$, separated by the energy $\delta_1$ of the anisotropic exchange interaction [7, 8, 30]. When a circularly polarized photon is absorbed, the superposition of the states $|X\rangle$ and $|Y\rangle$ is excited, the coherence of which (and hence the optical orientation signal) is lost within a characteristic time $\tau_c = \hbar/\delta_1$. The states $|X\rangle$ and $|Y\rangle$ are optically active in linear polarizations along the x and y axes, respectively, the absorption of linearly polarized light results in the optical alignment signal [31, 32]. The Faraday magnetic field $B > \delta_1/\mu_B g$ (where $\mu_B$ is the Bohr magneton, and the exciton g-factor is represented by the component of the Lande tensor $g_{zz}$) transforms the exciton fine structure into the circularly polarized states $|+1\rangle$ and $|-1\rangle$ (the Zeeman effect). Given that an exciton lifetime is sufficiently long $\tau > \hbar/\delta_1$, application of the magnetic field leads to the restoration of the optical orientation and suppression of the optical alignment [32]. The effects of conversion from the circular polarization of photoexcitation to the linear polarization of PL and vice versa in the magnetic field are also typical [32]. We note that unusual behavior of experimental dependences that is observed in the case of (In,Al)As QDs (see Sec. III and Discussion) requires more complex analysis.

To describe the experiment, we propose the two-step model of the exciton spin dynamics: an electron and a hole are excited by light in the $\Gamma$-valley of the Brillouin zone, followed by the electron energy relaxation into the X-valley, as shown in Fig. 1(a), while the heavy hole stays in the $\Gamma$-valley throughout its lifetime. Here, we use the pseudospin model, which does not take into account optically inactive exciton states, whereas the states of the optically active exciton $|+1\rangle$ and $|-1\rangle$ are associated with pseudospin projections $S_z = +1/2$ and $S_z = -1/2$, respectively [32]. The dipoles $|X\rangle$ and $|Y\rangle$ linearly polarized along the x and y axes (see Fig. 1(b)) are associated with projections $S_x = +1/2$ and $S_x = -1/2$, while the dipoles $|X'\rangle$ and $|Y'\rangle$ polarized along the x' and y' directions are described by projections $S_y = +1/2$ and $S_y = -1/2$. Note that the model is applicable in the low magnetic field limit $\delta_0 >> \mu_B g B$ and when $\delta_0 >> \delta_1$. In Discussion, the results of the pseudospin model are compared with the exact numerical solution obtained using the density matrix formalism.

Let us consider the first stage of the spin evolution: an exciton is photoexcited in the $\Gamma$-valley of the Brillouin zone and stays there during the time $\tau_\Gamma$ of the electron energy relaxation into the X-valley. For QDs elongated along the crystallographic direction [110], the stationary solution of the Bloch equation $d\boldsymbol{S}_\Gamma^{110}/dt = [\boldsymbol{\Omega}_\Gamma \times \boldsymbol{S}_\Gamma^{110}] - \boldsymbol{S}_\Gamma^{110}/T_S + \boldsymbol{S}_{0z}/\tau_\Gamma$ for average pseudospin under circularly polarized excitation is:

$$S_{\Gamma x}^{110} = \frac{S_{0z}T_s}{\tau_\Gamma}\frac{T_s{}^2\Omega_{EX,\Gamma}\Omega_{\|,\Gamma}}{1+T_s{}^2(\Omega_{EX,\Gamma}{}^2+\Omega_{\|,\Gamma}{}^2)}, \tag{3}$$

$$S_{\Gamma y}^{110} = \frac{S_{0z}T_s}{\tau_\Gamma}\frac{T_s\Omega_{EX,\Gamma}}{1+T_s{}^2(\Omega_{EX,\Gamma}{}^2+\Omega_{\|,\Gamma}{}^2)},$$

$$S_{\Gamma z}^{110} = \frac{S_{0z}T_s}{\tau_\Gamma}\frac{1+T_s{}^2\Omega_{\|,\Gamma}{}^2}{1+T_s{}^2(\Omega_{EX,\Gamma}{}^2+\Omega_{\|,\Gamma}{}^2)},$$

where $S_{0z}$ is the average pseudospin at the moment of exciton generation; $\boldsymbol{\Omega}_\Gamma = \boldsymbol{\Omega}_{\|,\Gamma} + \boldsymbol{\Omega}_{EX,\Gamma}$, $\Omega_{\|,\Gamma} = \mu_B g_\Gamma B/\hbar$, $\Omega_{EX,\Gamma} = \delta_{1\Gamma}/\hbar$, where $g_\Gamma$ and $\delta_{1\Gamma}$ are the g-factor and the anisotropic exchange splitting of the $\Gamma$-exciton states, respectively. The lifetime $T_s$ of the spin in the $\Gamma$-valley can be expressed through the spin relaxation time $\tau_s$ and the energy relaxation time of the electron into the X-valley as $T_s^{-1} = \tau_s^{-1} + \tau_\Gamma^{-1}$.

It will be shown that at the first stage (the $\Gamma$-valley), the limit $\Omega_\Gamma\tau_\Gamma \approx 1$ is realized, i.e. during a short stay in the excited state (about several ps), the exciton pseudospin has time to only partially rotate around the total field compounding of the exchange and external magnetic fields. Thus, in the X-state, all three components of the pseudospin will be non-zero, see Fig. 5. The indirect exciton lifetime in the ground state is relatively long (125 ns, see the Appendix section AI), as a result of which the limit $\Omega_X\tau_X >> 1$ is realized, where $\Omega_X = \sqrt{\Omega_{\|,X}^2 + \Omega_{EX,X}^2}$, $\Omega_{\|,X} = \mu_B g_X B/\hbar$ and $\Omega_{EX,X} = \delta_{1X}/\hbar$. To determine the average pseudospin components in the X-state, one should take the projection of the vector $\boldsymbol{S}_\Gamma$ onto the direction of the total field $(\Omega_{EX,X}, 0, \Omega_{\|,X})$, similar to how this was done in Refs. [33, 34]. As a result, for the optical orientation of excitons located in QDs elongated along the [110] axis, the following expression is obtained:

$$P_c^c = 2S_{Xz}^{110} = \frac{2S_{0z}T_s}{\tau_\Gamma}\frac{\Omega_{\|,X}(\Omega_{\|,X} + T_s{}^2\Omega_{EX,\Gamma}\Omega_{EX,X}\Omega_{\|,\Gamma} + T_s{}^2\Omega_{\|,X}\Omega_{\|,\Gamma}{}^2)}{(\Omega_{EX,X}{}^2+\Omega_{\|,X}{}^2)(1+T_s{}^2(\Omega_{EX,\Gamma}{}^2+\Omega_{\|,\Gamma}{}^2))}. \tag{4}$$

Note that the solution for QDs elongated along the [1-10] direction coincides with (4) up to the sign of $\Omega_{EX}$, which in this case does not affect the result, since (4) contains only terms quadratic in the exchange field. For QDs elongated along [100] (or [010]), the solution also coincides with (4), i.e. $S_{Xz}^{110} = S_{Xz}^{1-10} = S_{Xz}^{100} = S_{Xz}^{010}$. For the sake of model simplicity, we consider only the four specified directions along which the QD shape is distorted, it will be shown below that this is sufficient to describe the experiment.

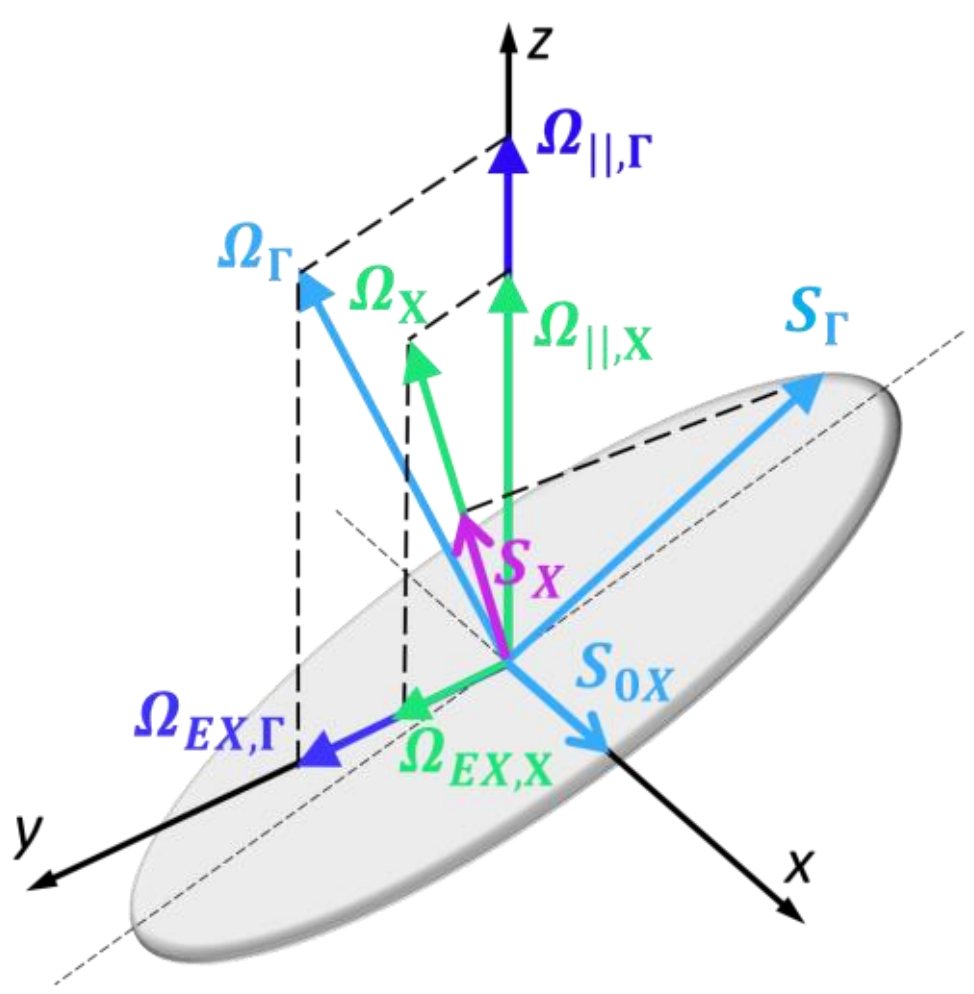

FIG. 5. The dynamics of pseudospin in the framework of the two-step process is shown for QDs elongated along the x' axis [100]. The pseudospin component $S_{0x}$ is excited by light linearly polarized along the x axis [110]. During the lifetime in the excited state ($\Gamma$-exciton), the pseudospin vector rotates by a finite angle around the direction $\mathbf{\Omega}_{\Gamma}$, which is a sum of the exchange $\mathbf{\Omega}_{EX,\Gamma}$ and Zeeman $\mathbf{\Omega}_{\parallel,\Gamma}$ components. To determine the average pseudospin components in the ground state (X-exciton), the projection onto the vector $\mathbf{\Omega}_{X}$ is taken.

Now let us consider the spin dynamics of the $\Gamma$-valley exciton for the QDs elongated along [110] with excitation linearly polarized along the [110] axis (x axis), which corresponds to pumping of the pseudospin component $S_{0x}$. The solution of the Bloch equation for the pseudospin is as follows:

$$S_{\Gamma x}^{110} = \frac{S_{0x}T_l}{\tau_{\Gamma}} \frac{1+T_l^2 \Omega_{EX,\Gamma}{}^2}{1+T_l^2(\Omega_{EX,\Gamma}{}^2 + \Omega_{\parallel,\Gamma}{}^2)}, \tag{5}$$

$$S_{\Gamma y}^{110} = \frac{S_{0x}T_l}{\tau_{\Gamma}} \frac{T_l \Omega_{\parallel,\Gamma}}{1+T_l^2(\Omega_{EX,\Gamma}{}^2 + \Omega_{\parallel,\Gamma}{}^2)},$$

$$S_{\Gamma z}^{110} = \frac{S_{0x}T_l}{\tau_{\Gamma}} \frac{T_l^2 \Omega_{EX,\Gamma} \Omega_{\parallel,\Gamma}}{1+T_l^2(\Omega_{EX,\Gamma}{}^2 + \Omega_{\parallel,\Gamma}{}^2)},$$

where $T_l^{-1} = \tau_l^{-1} + \tau_\Gamma^{-1}$, and $\tau_l$ is spin coherence time of an electron and a hole in an exciton. Projecting the vector $\boldsymbol{S}_\Gamma$ onto the direction $(\Omega_{EX,X}, 0, \Omega_{\parallel,X})$ yields the pseudospin components in the X-state ($\boldsymbol{S}_X$). The optical alignment signal is expressed as:

$$P_l^{l_[110]} = 2S_{Xx}^{110} = \frac{2S_{0x}T_l}{\tau_\Gamma} \frac{\Omega_{EX,X}(\Omega_{EX,X} + T_l^2\Omega_{EX,X}\Omega_{EX,\Gamma}{}^2 + T_l^2\Omega_{EX,\Gamma}\Omega_{\parallel,\Gamma}\Omega_{\parallel,X})}{(\Omega_{EX,X}{}^2 + \Omega_{\parallel,X}{}^2)(1 + T_l^2(\Omega_{EX,\Gamma}{}^2 + \Omega_{\parallel,\Gamma}{}^2))}. \quad (6)$$

For QDs elongated along [1-10], the solution coincides with (6): $S_{Xx}^{110} = S_{Xx}^{1-10}$. For QDs elongated along [100] (or [010]), the optical alignment signal $P_l^{l_[100]}$ is absent. This is due to the fact that the projection $\boldsymbol{S}_\Gamma$ in the case of [100] QDs is taken onto the direction $(0, \Omega_{EX,X}, \Omega_{\parallel,X})$ and $P_l^{l_[100]} = 2S_{Xx}^{110}$ is identically zero.

The conversion signal from linear polarization [110] to circular for QDs elongated along the [110] axis is described by the expression:

$$P_c^{l_[110]} = 2S_{Xz}^{110} = \frac{2S_{0x}T_l}{\tau_\Gamma} \frac{\Omega_{\parallel,X}(\Omega_{EX,X} + T_l^2\Omega_{EX,X}\Omega_{EX,\Gamma}{}^2 + T_l^2\Omega_{EX,\Gamma}\Omega_{\parallel,\Gamma}\Omega_{\parallel,X})}{(\Omega_{EX,X}{}^2 + \Omega_{\parallel,X}{}^2)(1 + T_l^2(\Omega_{EX,\Gamma}{}^2 + \Omega_{\parallel,\Gamma}{}^2))}. \quad (7)$$

For [1-10] QDs the solution changes sign: $S_{Xz}^{110} = -S_{Xz}^{1-10}$.

The conversion signal from linear [110] to circular polarization for QDs elongated along [100] is described by the even dependence:

$$P_c^{l_[100]} = 2S_{Xz}^{100} = -2S_{Xz}^{010} = \frac{-2S_{0x}T_l}{\tau_\Gamma} \frac{T_l\Omega_{\parallel,X}(\Omega_{EX,\Gamma}\Omega_{\parallel,X} - \Omega_{EX,X}\Omega_{\parallel,\Gamma})}{(\Omega_{EX,X}{}^2 + \Omega_{\parallel,X}{}^2)(1 + T_l^2(\Omega_{EX,\Gamma}{}^2 + \Omega_{\parallel,\Gamma}{}^2))}. \quad (8)$$

When QDs elongated along [110] are excited by light linearly polarized along [100], the rotation of the linear polarization plane in the magnetic field takes place, which is described by the relation:

$$P_l^{l'_[110]} = 2S_{Xx}^{110} = 2S_{Xx}^{1-10} = \frac{2S_{0x}T_l^2}{\tau_\Gamma} \frac{\Omega_{EX,X}(\Omega_{EX,X}\Omega_{\parallel,\Gamma} - \Omega_{EX,\Gamma}\Omega_{\parallel,X})}{(\Omega_{EX,X}{}^2 + \Omega_{\parallel,X}{}^2)(1 + T_l^2(\Omega_{EX,\Gamma}{}^2 + \Omega_{\parallel,\Gamma}{}^2))}. \quad (9)$$

The linear polarization rotation signal for [100] and [010] QDs is identically zero: $P_l^{l'_[100]} = P_l^{l'_[010]} \equiv 0$.

### B. Two-step spin evolution model within the density matrix formalism

Calculations of the two-step spin dynamics were performed within the density matrix formalism as well. In addition to the exchange interaction the hyperfine interaction of the electrons with nuclei was taken into account. We make the standard assumption that the characteristic time of the random nuclear field variation exceeds the exciton lifetime, meaning that the nuclear field is "frozen" [35]. The nuclear field $\boldsymbol{B}_N$ dispersion in a QDs ensemble is described by a three-dimensional normal distribution:

$$W(\boldsymbol{B}_N) = \frac{1}{\pi^{3/2}\Delta_B^3}\exp\left[-\frac{\boldsymbol{B}_N^2}{\Delta_B^2}\right]. \tag{10}$$

Since the nuclear field directed in the quantum dot plane mixes the $\left|\pm1\right\rangle$ and $\left|\pm2\right\rangle$ states of direct and indirect excitons, the consideration within the framework of the pseudospin formalism is no longer valid and it is necessary to solve the stationary Lindblad equation for the density matrix of the entire system:

$$0 = -\frac{i}{\hbar}[H,\rho] + \sum_k \gamma_k \left( L_k \rho L_k^\dagger - \frac{1}{2}\left\{L_k^\dagger L_k, \rho\right\}\right), \tag{11}$$

where $L_k$ are the jump operators, $\gamma_k$ are the jump rates. We write the Hamiltonian $H$ and the density matrix $\rho$ in the basis of exciton states $\left|+2_\Gamma, -1_\Gamma, +1_\Gamma, -2_\Gamma, +2_X, -1_X, +1_X, -2_X\right\rangle$. The Hamiltonian of the full system has a block form:

$$H = \begin{bmatrix} H_\Gamma & H_{\text{mix}} \\ H_{\text{mix}}{}^* & H_X \end{bmatrix}. \tag{12}$$

The diagonal contains blocks for the direct exciton $H_\Gamma$ (Eq. (A1)) and the indirect exciton $H_X$ (Eq. (A2)). The $H_\Gamma$ block for the direct exciton includes the isotropic and anisotropic exchange interactions, the external magnetic field and the random field of the nuclei. The $H_X$ block for the indirect exciton includes only the external magnetic field and the random field of the nuclei, since the electron and the hole are at the different points in momentum space, and the exchange interaction is strongly suppressed. The off-diagonal blocks consist of diagonal matrices $H_{\text{mix}}$ (Eq. (A3)), which are responsible for the $\Gamma$-X mixing of direct and indirect exciton states [36].

For a direct exciton, the hopping operators contain contributions from the radiative recombination of $\left|\pm1_\Gamma\right\rangle$ states $\gamma_{r,\pm1\Gamma}$, as well as the energy relaxation $\gamma_{\Gamma\text{X}} = 1/\tau_\Gamma$ into the indirect exciton $\left|\pm1_X\right\rangle$. It is assumed that during $\Gamma$-X relaxation from a direct exciton to an indirect one, the spins of the electron and the hole are conserved. The spin relaxation of the electron and the hole not associated with either exchange or hyperfine interaction is presumed to be infinitely long, which is consistent with the microsecond spin relaxation times measured experimentally in this

system [12]. Therefore, we do not introduce the jump operators responsible for additional spin relaxation of electrons and holes.

We verified that, instead of calculating for the full matrix, one can take into account $\Gamma$-X mixing by adding nonzero isotropic and anisotropic exchange splittings in the indirect exciton block $H_X$ (Eq. (A5)). In this case, the Hamiltonian matrix becomes block-diagonal (Eq. (A9)), and the interaction between the direct and indirect exciton states is contained only in the incoherent part of the Lindblad equation, which describes the energy relaxation of an electron from the $\Gamma$-valley to the X-valley. In addition, due to the $\Gamma$-X mixing, the states of the indirect exciton $\left|\pm 1_\Gamma\right\rangle$ acquire a finite radiative recombination lifetime $\tau_X{}^{-1} = \left(V_{\mathrm{mix}} / \Delta E_{\Gamma \mathrm{X}}\right)^2 \tau_{r\Gamma}{}^{-1}$, where $V_{\mathrm{mix}}$ is the matrix element of the coupling between the $\left|\Gamma\right\rangle$ and $\left|\mathrm{X}\right\rangle$ states, $\Delta E_{\Gamma\mathrm{X}}$ is the energy splitting of the $\left|\Gamma\right\rangle$ and $\left|\mathrm{X}\right\rangle$ exciton states.

The calculation of the PL polarization was performed by averaging over 100 quantum dots, with the nuclear field described by the distribution $W(\boldsymbol{B}_N)$. We neglect possible anisotropy of the hyperfine interaction [16] in the X-valley and assume the same magnitude of the nuclear fields for electrons in the $\Gamma$ and X-valleys. The exact value of the nuclear field in the $\Gamma$-valley, which can be several times larger [16] is not important for the modeling. Due to the large anisotropic splitting of the direct exciton and fast relaxation time $\tau_\Gamma$, hyperfine interaction does not affect the exciton spin dynamics.

## V. DISCUSSION

The experimental data shown in Fig. 3(a,b) and 4(a,b) were modeled within the pseudospin, Eq. (4, 6-9), and density matrix, Eq. (11,12), methods. The model curves within the pseudospin formalism and density matrix formalism are shown in the figures by the dashed red lines and solid green lines, respectively. A good agreement with the experimental data is achieved with the fitting parameters of both models given in Table 1. For the modelling we take the electron and heavy hole g-factor values ($g_{zz}$) determined in Refs. [36, 37]. The same anisotropic exchange splitting of the direct exciton states of 210 μeV is used in both methods. This value is consistent with the value of 260 μeV for (In,Al)As/AlAs QDs obtained in Ref. [11] and with values in Ref. [18], as well as with the typical splitting of direct exciton states in other systems [6-8]. The anisotropic splitting $\delta_{1X}$ of the indirect exciton determined using the pseudospin model is 2.6 times larger as compared to the value determined within the density matrix model. The reason is that in the density matrix approach the spin dynamics of an X-exciton is controlled by the joint effect of the anisotropic exchange splitting and the hyperfine interaction. From the modelling, the average field of the nuclear fluctuations $B_N = \sqrt{3/2}\Delta_B = 6.1$ mT was obtained, what agrees well with the value of 1.25 mT from Ref. [16]. The electron thermalization times from the $\Gamma$ to the X-valley of 4.7 and 3.8 ps determined within the two methods are consistent with picosecond relaxation times measured in Ref. [24].

| | $\delta_{0\Gamma}$ (μeV) | $\delta_{1\Gamma}$ (μeV) | $\delta_{0X}$ (μeV) | $\delta_{1X}$ (μeV) | $\tau_\Gamma$ (ps) | $\Delta E_{\Gamma X}$ (meV) | $B_N$ (mT) | $\frac{N_{110}}{N_{100}}$ | $g_{e\Gamma}$ | $g_{eX}$ | $g_h$ |
|---|---|---|---|---|---|---|---|---|---|---|---|
| Pseudospin model | - | 210 | - | 1.3 | 4.7 | - | - | 3/1 | -0.2 | 2 | 2.43 |
| Lindblad Equation | 1000 | 210 | 1.6 | 0.7 | 3.8 | 22.5 | 6.1 | 4/1 | -0.2 | 2 | 2.43 |

Table 1. Parameters used for modeling of the experimental data within the pseudospin and Lindblad equation models: $\delta_{0\Gamma,X}$ – isotropic exchange splitting for the direct and indirect exciton states; $\delta_{1\Gamma,X}$ – anisotropic exchange splitting for the direct and indirect exciton states; $\tau_\Gamma$ – electron relaxation time from the $\Gamma$ to X-valley; $\Delta E_{\Gamma X}$ – energy splitting of the $|\Gamma\rangle$ and $|X\rangle$ exciton states; $B_N$ – the average nuclear field; $N_{100}, N_{110}$ – relative number of QDs elongated along the [100] and [110] crystallographic axes, respectively; $g_{e\Gamma}$ – g-factor of an electron in the $\Gamma$-valley; $g_{eX}$ – g-factor of an electron in the X-valley; $g_h$ – longitudinal g-factor of a heavy hole in the $\Gamma$-valley.

Now let us discuss what determines the peculiarities of the experimental dependencies. The two-component recovery curve of the optical orientation shown in Fig. 3(a) is caused by a partial loss (from 81 to 35%) of the spin polarization in the intermediate $\Gamma$-state of an exciton followed by the depolarization in the X-state (from 35 to 11%). This becomes possible due to comparable time of the electron relaxation between $\Gamma$ and X-valleys and the spin precession period in the anisotropic exchange field, within the peudospin model we obtain $\tau_\Gamma = 4.7\,\text{ps} \sim 1/\Omega_{EX,\Gamma} = 3.14$ ps. The external magnetic field exceeding the exchange field $B_{EX,\Gamma} = \delta_{1\Gamma}/\mu_B g_\Gamma = 1.4$ T is required to restore the spin polarization of the direct exciton. For the indirect exciton, application of the magnetic field $\delta_{1\Gamma}/\mu_B g_\Gamma > B > \delta_{1X}/\mu_B g_X$ (about 50 mT) allows spin relaxation in the X-valley to be suppressed and the optical orientation to be restored from 11 to 35%. We note that the magnetically induced circular polarization under the given excitation conditions turns out to be negligible, and the optical orientation recovery curve is almost symmetrical with respect to the sign of the field. In addition, dynamic polarization of electrons [14] could be observed, leading to a violation of the curve parity in the range of small fields, but in this case the effect turns out to be negligible as well.

The indirect exciton lifetime of 125 ns (see Appendix) is several orders of magnitude larger than the spin precession period 0.5 ns in the anisotropic exchange field. It means that optical orientation should be completely lost during the exciton lifetime, what contradicts to the residual circular polarization of 11% observed in the experiment at zero external magnetic field. Within the pseudospin model, the existence of a finite optical orientation in zero magnetic field requires the assumption of an additional subensemble of QDs lacking anisotropic exchange splitting that contribute to the PL signal. These could be symmetric neutral quantum dots, in which there is no splitting of the exciton states $\left|\pm1\right\rangle$ into linearly polarized states $\left|X\right\rangle$ and $\left|Y\right\rangle$. These could also be charged quantum dots, in which a trion is formed upon optical excitation [38]. Within the density matrix approach, the finite optical orientation signal in zero field is shown to be a consequence of competition between the anisotropic exchange interaction in the exciton and the hyperfine interaction of the electron with nuclei in the same ensemble of neutral QDs. The fitting shows that in the studied QDs, the anisotropic exchange splitting of the indirect exciton states and the characteristic splitting of the X-valley electron states in the nuclear random field are approximately equal: $\delta_{1X} = 0.7$ μeV and $\mu_B g_{eX} \Delta_B = 0.71$ μeV. Thus, the QDs are in an intermediate regime between two limiting cases. In the first limiting case, the anisotropic exchange splitting dominates, which would lead to the absence of optical orientation of the indirect exciton in zero magnetic field and to polarization recovery in low fields. In the second limiting case, when the exchange filed is negligible compared to a nuclear field, one would observe only a broad contour (1.4 T) from the optical orientation recovery of the direct exciton. This is due to the fact that the electron-nuclear hyperfine interaction does not couple the $\left|\pm1\right\rangle$ states and cannot suppress the optical orientation since the microsecond time scales of the hole spin relaxation in (In,Al)As/AlAs QDs [12], the hyperfine interaction of the holes with nuclei is neglected.

In contrast to the optical orientation, optical alignment suppression in the external magnetic field is a one-step process (see the one-component curve in Fig. 3(b)). The reason is the anisotropic exchange splitting suppression of the indirect exciton states in small magnetic fields (50 mT), so that $\left|\pm1_X\right\rangle$ states split in the external field become the eigenstates. As a result, coherence of these states corresponding to linearly polarized states $\left|X_X\right\rangle$ and $\left|Y_X\right\rangle$, is destroyed during the indirect exciton lifetime. As a result, the exciton alignment transferred from the direct exciton is lost.

The model dependence of the conversion from linear to circular polarization (Fig. 4(a)) also contains two components corresponding to the spin dynamics of direct and indirect excitons. However, it is not possible to reliably establish the presence of a narrow contour in the experimental data. The relatively small amplitude of the conversion effect (10%) as compared to the amplitude of the suppression of optical alignment (35%) is due to the presence of QDs elongated along different crystallographic directions, for simplicity we consider the four directions: [110], [1-10], [100], [010]. When the number of QDs elongated along [110] is exactly equal to the number of QDs elongated along [1-10] (and similarly for the [100] and [010] axes), the amplitude of the conversion effect will be exactly zero. It should be noted, that the optical orientation and optical alignment dependencies are not sensitive to the direction of the QD in-plane anisotropy axis.

For fitting of the linear-to-circular conversion signal we use the ratios N[110]/N[100] = 3/1 and 4/1 in the pseudospin model and the density matrix model, respectively. Here N[110] stands for the number of QDs elongated along the [110] or [1-10] axis and N[100] stands for the number of [100] or [010] QDs. These ratios are consistent with the relation $P_l^l / P_{l'}^{l'} \approx 3/1$ of the optical alignment signal magnitude, observed experimentally in Ref. [19]. Thus, to describe the conversion signal in the pseudospin model we used the following relation for PL polarization:

$$P_c^l = \tfrac{3}{4}(P_c^{l_[110]} + P_c^{l_[1-10]}) + \tfrac{1}{4}(P_c^{l_[100]} + P_c^{l_[010]}). \quad (13)$$

The center of gravity of the conversion signal in Fig. 4(a) is shifted to the negative values. According to our modeling, the reason is different magnetic field dependence of the conversion signal from QDs elongated along the [110]/[1-10] and [100]/[010] axes. For the QDs, whose anisotropy axis coincide with the laser polarization plane, the conversion signal shows typical asymmetric behavior, see Eq. (7). For QDs, whose anisotropy axis is tilted at 45 degrees from the laser polarization plane, the conversion signal shows even behavior, see Eq. (8), which, as far as we know, has never been reported before.

For rotation of the linear polarization plane shown in Fig. 4(b) we again observe a one-step process, as in the case of the optical alignment suppression. Noticeably, that the conversion from linear polarization in the dashed axes to polarization in the "main" axes and vice versa should be absent if the indirect excitons would be directly created, since the condition $\Omega_{EX,X}\tau_X >> 1$ ($\tau_X >> \hbar/\delta_{1X}$) is fulfilled. The conversion becomes possible due to initial excitation of direct excitons. Before the $\Gamma$-X relaxation event, the direct exciton pseudospin makes incomplete rotation around the effective magnetic field axis formed by the anisotropic exchange field and external magnetic field (see Fig. 5). As a result, after the electron relaxation to the X-valley, all of the pseudospin projections are non-zero.

In the modeling within the density matrix approach we took into account the dispersion of the $\Gamma$-X mixing parameter across the QD ensemble described by the normal distribution with a mean value of $V_{\text{mix}} = 1.3$ meV and a standard deviation of 0.1 meV associated with the variations of sharpness of QDs heteroboundaries. We introduced the dispersion of $V_{\text{mix}}$ because the presence of transverse nuclear spin fluctuations lead to the anti-crossing of the $\left|\pm 1_X\right\rangle$ and $\left|\pm 2_X\right\rangle$ states, what should manifest itself as resonance peaks in the optical orientation and optical alignment effects at low external magnetic fields about 25 mT. The absence of these features in the experiment indicates a spread in the $\delta_{0X}$ and $\delta_{1X}$ values across the QDs ensemble, leading to a broadening of the resonance, which we considered in our calculations. The absence of this resonance may also be due to a spread in the exciton g-factor in the ensemble.

## VI. CONSCLUSIONS

The exciton optical orientation, optical alignment, linear-to-circular conversion effect and linear polarization plane rotation were studied in type-I indirect (In,Al)As/AlAs QDs subject to magnetic field. The peculiarities of the effects were attributed to the two-step exciton lifecycle: generation of the direct state, followed by energy relaxation in the indirect state. This allows one to effectively excite the indirect in momentum space excitons with a weak oscillator strength which have extremely long lifetimes and spin relaxation times. We showed that the spin relaxation in the intermediate direct state can affect the exciton spin dynamics. Description within the pseudospin and the density matrix models allowed to determine parameters, such as anisotropic exchange splittings of the direct and indirect exciton states, the last one is comparable in magnitude with the splitting of the electron states caused by the hyperfine interaction with nuclei (about 1 μeV or less). We note that the specified feature of the fine structure of indirect in momentum space excitons should lead to new features in the PL polarization signal in a small magnetic field what requires further research of this uncommon model system.

## ACKNOWLEDGMENTS

The authors are grateful to D. S. Smirnov for fruitful discussions. The development of the theoretical model within the pseudospin formalism and the experimental studies, including measurements of resonant PL polarization subject to magnetic field conducted by S.V.N., M.D.R., Ya.A.K. and Yu.G.K. were supported by the Russian Science Foundation (Grant No. 22-12-00125-P). Sample preparation and sample characterization conducted by T.S.S. was supported by the Russian Science Foundation (Grant No. 22-12-00022-P). Theoretical model within the density matrix formalism and the time-resolved PL measurements carried out by A.A.G., I.V.K. and N.O.M. were supported by the Ministry of Science and Higher Education of the Russian Federation (project FFUG-2024-0037).

[1] X. Marie, B. Urbaszek, O. Krebs, T. Amand, Exciton spin dynamics in semiconductor quantum dots, Spin Physics in Semiconductors, edited by M. I. Dyakonov, Springer, Berlin, 2008.

[2] A. Chatterjee, P. Stevenson, S. De Franceschi, A. Morello, N. P. de Leon, F. Kuemmeth, Semiconductor qubits in practice, Nat. Rev. Phys. 3 (2021) 157–177. https://doi.org/10.1038/s42254-021-00283-9

[3] G. Burkard, T. D. Ladd, A. Pan, J. M. Nichol, J. R. Petta, Semiconductor spin qubits, Rev. Mod. Phys. 95 (2023) 025003. https://doi.org/10.1103/RevModPhys.95.025003

[4] A. S. Bracker, D. Gammon, V. L. Korenev, Fine structure and optical pumping of spins in individual semiconductor quantum dots, Semicond. Sci. Technol. 23 (2008) 114004. DOI 10.1088/0268-1242/23/11/114004

[5] A. V. Khaetskii, Yu. V. Nazarov, Spin relaxation in semiconductor quantum dots, Phys. Rev. B 61 (2000) 12639. https://doi.org/10.1103/PhysRevB.61.12639

[6] M. Paillard, X. Marie, P. Renucci, T. Amand, A. Jbeli, J.M. Gérard, Spin relaxation quenching in semiconductor quantum dots, Phys. Rev. Lett. 86 (2001) 1634-1637. ttps://doi.org/10.1103/PhysRevLett.86.1634

[7] D. Gammon, E. S. Snow, B. V. Shanabrook, D. S. Katzer, D. Park, Fine structure splitting in the optical spectra of single GaAs quantum dots, Phys. Rev. Lett. 76 (1996) 3005-3008. https://doi.org/10.1103/PhysRevLett.76.3005

[8] M. Bayer, G. Ortner, O. Stern, A. Kuther, A. A. Gorbunov, A. Forchel, P. Hawrylak, S. Fafard, K. Hinzer, T. L. Reinecke, S. N. Walck, J. P. Reithmaier, F. Klopf, F. Schäfer, Fine structure of neutral and charged excitons in self-assembled In(Ga)As/(Al)GaAs quantum dots, Phys. Rev. B 65 (2002) 195315. https://doi.org/10.1103/PhysRevB.65.195315

[9] H. Yu, S. Lycett, C. Roberts, R. Murray, Time resolved study of self-assembled InAs quantum dots, Appl. Phys. Lett. 69 (1996) 4087-4089. https://doi.org/10.1063/1.117827

[10] T. S. Shamirzaev, A. M. Gilinsky, A. K. Bakarov, A. I. Toropov, D. A. Ténné, K. S. Zhuravlev, C. von Borczyskowski, D. R. T. Zahn, Millisecond photoluminescence kinetics in a system of direct-bandgap InAs quantum dots in an AlAs matrix, JEPT Letters 77 (2003) 389-392. https://doi.org/10.1134/1.1581967

[11] J. Rautert, T. S. Shamirzaev, S. V. Nekrasov, D. R. Yakovlev, P. Klenovský, Yu. G. Kusrayev, M. Bayer, Optical orientation and alignment of excitons in direct and indirect band gap

(In,Al)As/AlAs quantum dots with type-I band alignment, Phys. Rev. B 99 (2019) 195411. https://doi.org/10.1103/PhysRevB.99.195411
[12] T. S. Shamirzaev, D. R. Yakovlev, D. S. Smirnov, V. N. Mantsevich, M. Bayer, Scaling laws of electron and hole spin relaxation in indirect band gap (In,Al)As/AlAs quantum dots, Phys. Rev. B 114 (2026) 115415. DOI: https://doi.org/10.1103/b358-1776
[13] T. S. Shamirzaev, D. R. Yakovlev, V. N. Mantsevich, D. Kudlacik, A. Yu. Gornov, A. K. Gutakovskii, M. Bayer, Magnetic field induced exciton spin dynamics in indirect band gap (In,Al)As/AlAs quantum dots, J. Lumin. 288 (2025) 121596. https://doi.org/10.1016/j.jlumin.2025.121596
[14] D. S. Smirnov, T. S. Shamirzaev, D. R. Yakovlev, M. Bayer, Dynamic polarization of electron spins interacting with nuclei in semiconductor nanostructures, Phys. Rev. Lett. 125 (2020) 156801. https://doi.org/10.1103/PhysRevLett.125.156801
[15] T. S. Shamirzaev, A. V. Shumilin, D. S. Smirnov, D. Kudlacik, S. V. Nekrasov, Yu. G. Kusrayev, D. R. Yakovlev, M. Bayer, Optical orientation of excitons in a longitudinal magnetic field in indirect-band-gap (In,Al)As/AlAs quantum dots with type-I band alignment, Nanomaterials 13 (2023) 729. https://doi.org/10.3390/nano13040729
[16] M. S. Kuznetsova, J. Rautert, K. V. Kavokin, D. S. Smirnov, D. R. Yakovlev, A. K. Bakarov, A. K. Gutakovskii, T. S. Shamirzaev, M. Bayer, Electron-nuclei interaction in the X valley of (In,Al)As/AlAs quantum dots, Phys. Rev. B 101 (2020) 075412. https://doi.org/10.1103/PhysRevB.101.075412
[17] T. S. Shamirzaev, A. V. Nenashev, A. K. Gutakovskii, A. K. Kalagin, K. S. Zhuravlev, M. Larsson, P. O. Holtz, Atomic and energy structure of InAs/AlAs quantum dots, Phys. Rev. B 78 (2008) 085323. https://doi.org/10.1103/PhysRevB.78.085323
[18] J. Rautert, M. V. Rakhlin, K. G. Belyaev, T. S. Shamirzaev, A. K. Bakarov, A. A. Toropov, I. S. Mukhin, D. R. Yakovlev, M. Bayer, Anisotropic exchange splitting of excitons affected by mixing in (In,Al)As/AlAs quantum dots: Microphotoluminescence and macrophotoluminescence measurements, Phys. Rev. B 100 (2019) 205303. https://doi.org/10.1103/PhysRevB.100.205303
[19] S. V. Nekrasov, N. O. Mikhailenko, M. D. Ragoza, T. S. Shamirzaev, Yu. G. Kusrayev, Influence of Γ-X mixing on optical orientation and alignment of excitons in (In,Al)As/AlAs quantum dots, Phys. Rev. B 110 (2024) 115435. https://doi.org/10.1103/PhysRevB.110.115435
[20] D. S. Smirnov, E. L. Ivchenko, Theory of polarized photoluminescence of indirect band gap excitons in type-I quantum dots, Phys. Rev. B 108 (2023) 195432. https://doi.org/10.1103/PhysRevB.108.195432

[21] D. S. Smirnov, E. L. Ivchenko. Interplay between hyperfine and anisotropic exchange interactions in exciton luminescence of quantum dots. Opt. Spectr. 132 (2024) 864-868. https://doi.org/10.61011/OS.2024.08.59033.6714-24
[22] O. Benson, C. Santori, M. Pelton, Y. Yamamoto, Regulated and entangled photons from a single quantum dot. Phys. Rev. Lett. 84 (2000) 2513–2516. https://doi.org/10.1103/PhysRevLett.84.2513
[23] C. Santori, D. Fattal, M. Pelton, G. S. Solomon, Y. Yamamoto, Polarization-correlated photon pairs from a single quantum dot. Phys. Rev. B 66 (2002) 045308. https://doi.org/10.1103/PhysRevB.66.045308
[24] T. S. Shamirzaev, D. S. Abramkin, A. V. Nenashev, K. S. Zhuravlev, F. Trojanek, B. Dzurnak, P. Maly, Carrier dynamics in InAs/AlAs quantum dots: lack in carrier transfer from wetting layer to quantum dots, Nanotechnology 21 (2010) 155703-155707. DOI:10.1088/0957-4484/21/15/155703
[25] T. S. Shamirzaev, D. S. Abramkin, D. V. Dmitriev, A. K. Gutakovskii, Nonradiative energy transfer between vertically coupled indirect and direct bandgap InAs quantum dots, Appl. Phys. Lett. 97 (2010) 263102. https://doi.org/10.1063/1.3532102
[26] T. S. Shamirzaev, A. M. Gilinsky, A. K. Kalagin, A. I. Toropov, A. K. Gutakovskii, K. S. Zhuravlev, Strong sensitivity of photoluminescence of InAs/AlAs quantum dots to defects: Evidence for lateral inter-dot transport, Semicond. Sci. Technol. 21 (2006) 527. DOI:10.1088/0268-1242/21/4/019
[27] M. G. Bawendi, W. L. Wilson, L. Rothberg, P. J. Carroll, T.M. Jedju, M. L. Streigerwald, L. E. Brus, Electronic structure and photoexcited-carrier dynamics in nanometer-size CdSe clusters, Phys. Rev. Lett. 65 (1990) 1623-1626. https://doi.org/10.1103/PhysRevLett.65.1623
[28] M. Nirma, C. B. Murray, M. G. Bawendi, Fluorescence-line narrowing in CdSe quantum dots: Surface localization of the photogenerated exciton, Phys. Rev. B 50 (1994) 2293-2300. https://doi.org/10.1103/PhysRevB.50.2293
[29] O. Madelung, M. Schulz, H. Weiss, Landolt‒Bornstein, Semiconductors: other than group IV elements and III–V compounds. – Springer Science & Business Media, 2012.
[30] E. L. Ivchenko, G. E. Pikus, Superlattices and Other Heterostructures: Symmetry and Optical phenomena, Springer, Berlin, 1997.
[31] G. E. Pikus, E. L. Ivchenko, Excitons, edited by E. I. Rashba, M. D. Struge, North-Holland, Amsterdam, 1982.
[32] R. I. Dzhioev, H. M. Gibbs, E. L. Ivchenko, G. Khitrova, V. L. Korenev, M. N. Tkachuk, B. P. Zakharchenya, Determination of interface preference by observation of linear-to-circular

polarization conversion under optical orientation of excitons in type-II GaAs/AlAs superlattices, Phys. Rev. B 56 (1997) 13405-13413. https://doi.org/10.1103/PhysRevB.56.13405
[33] Yu. G. Kusrayev, A. V. Koudinov, B. P. Zakharchenya, S. Lee, J. K. Furdyna, M. Dobrowolska, Optical orientation and alignment of excitons in self-assembled CdSe/ZnSe quantum dots: The role of excited states, Phys. Rev. B 72 (2005) 155301. https://doi.org/10.1103/PhysRevB.72.155301
[34] A. V. Koudinov, B. R. Namozov, Yu. G. Kusrayev, S. Lee, M. Dobrowolska J. K. Furdyna. Two-step versus one-step model of the interpolarization conversion and statistics of CdSe/ZnSe quantum dot elongations, Phys. Rev. B 78 (2008) 045309. https://doi.org/10.1103/PhysRevB.78.045309
[35] I. A. Merkulov, Al. L. Efros, M. Rosen, Electron spin relaxation by nuclei in semiconductor quantum dots, Phys. Rev. B 65 (2002) 205309. https://doi.org/10.1103/PhysRevB.65.205309
[36] J. Debus, T. S. Shamirzaev, D. Dunker, V. F. Sapega, E. L. Ivchenko, D. R. Yakovlev, A. I. Toropov, M. Bayer, Spinflip Raman scattering of the Γ-X mixed exciton in indirect band gap (In,Al)As/AlAs quantum dots, Phys. Rev. B 90 (2014) 125431. https://doi.org/10.1103/PhysRevB.90.125431
[37] I. A. Yugova, A. Greilich, D. R. Yakovlev, A. A. Kiselev, M. Bayer, V. V. Petrov, Yu. K. Dolgikh, D. Reuter, A. D. Wieck, Universal behavior of the electron *g* factor in GaAs/Al*x*Ga1−*x*As quantum wells, Phys. Rev. B 75 (2007) 245302. https://doi.org/10.1103/PhysRevB.75.245302
[38] A. S. Bracker, E. A. Stinaff, D. Gammon, M. E. Ware, J. G. Tischler, A. Shabaev, Al. L. Efros, D. Park, D. Gershoni, V. L. Korenev, I. A. Merkulov, Optical Pumping of the Electronic and Nuclear Spin of Single Charge-Tunable Quantum Dots, Phys. Rev. Lett. 94 (2005) 047402. https://doi.org/10.1103/PhysRevLett.94.047402

# APPENDIX

## AI. Indirect exciton lifetime measurement

In order to determine the lifetime of the indirect in momentum space exciton localized in a (In,Al)As/AlAs QD time-resolved measurements were performed. For this purpose, an electro-optic modulator (EOM) operating in a transmission/blocking mode at a frequency of 100 kHz was inserted into the PL excitation path. The switching time of the modulator is 13 ns. From the rise of the PL signal after opening the EOM, see Fig. A(a), the lifetime of the indirect exciton was determined to be 125 ns, in agreement with the value obtained from the PL decay following the closure of the EOM.

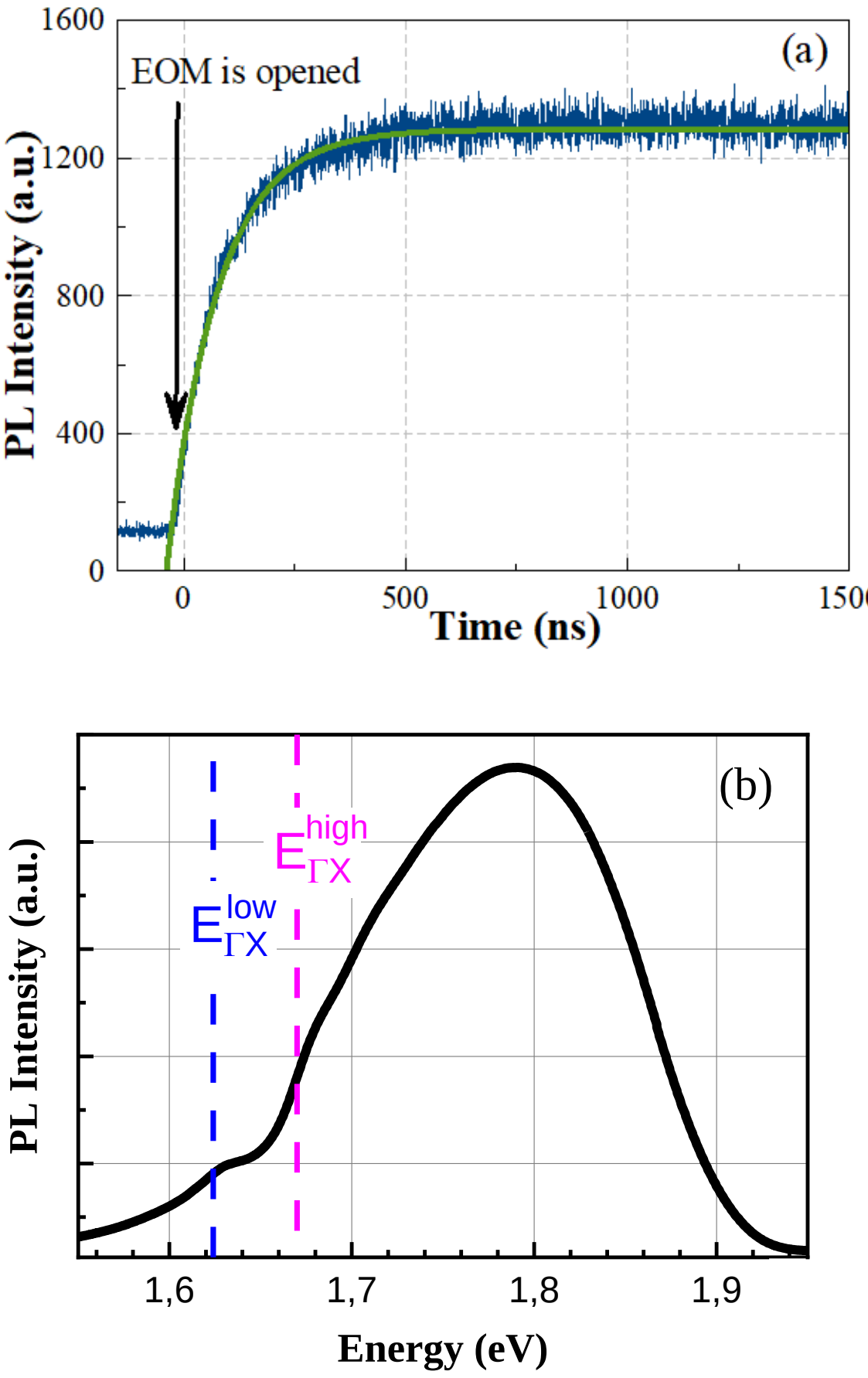

Figure A. (a) Time dependence of the PL intensity rise after the opening of the laser pumping. The experimental data (shown in blue) are described by an exponential function (solid green line). The PL excitation energy is 1.66 eV, the detection energy is 1.64 eV. (b) PL spectrum measured at the nonresonant excitation energy of 3.07 eV [19]. $E_{\Gamma X}^{low}$ and $E_{\Gamma X}^{high}$ stand for the $\Gamma$-X crossing energies, for details see Ref. [19]. The temperature is 14 K.

## AII. Density matrix formalism details

The matrix form of the direct exciton block $H_\Gamma$:

$$H_\Gamma = \frac{1}{2}\begin{bmatrix} -\delta_{0\Gamma} & 0 & 0 & 0 \\ 0 & \delta_{0\Gamma} & \delta_{1\Gamma} & 0 \\ 0 & \delta_{1\Gamma} & \delta_{0\Gamma} & 0 \\ 0 & 0 & 0 & -\delta_{0\Gamma} \end{bmatrix} + \frac{\mu_B B}{2}\begin{bmatrix} -g_{\Gamma,2} & 0 & 0 & 0 \\ 0 & -g_\Gamma & 0 & 0 \\ 0 & 0 & g_\Gamma & 0 \\ 0 & 0 & 0 & g_{\Gamma,2} \end{bmatrix}$$

$$+\frac{\mu_B g_{e\Gamma}}{2}\begin{bmatrix} B_{N\|}{}^{\Gamma} & 0 & B_{N\perp}{}^{\Gamma} & 0 \\ 0 & B_{N\|}{}^{\Gamma} & 0 & B_{N\perp}{}^{\Gamma} \\ B_{N\perp}{}^{\Gamma*} & 0 & -B_{N\|}{}^{\Gamma} & 0 \\ 0 & B_{N\perp}{}^{\Gamma*} & 0 & -B_{N\|}{}^{\Gamma} \end{bmatrix}, \tag{A1}$$

here $\delta_{0\Gamma}$ is the isotropic exchange splitting of the $|\pm1_{\Gamma}\rangle$ and $|\pm2_{\Gamma}\rangle$ states of the direct exciton, $\delta_{1\Gamma}$ is the anisotropic exchange splitting of the $|\pm1_{\Gamma}\rangle$ states of the direct exciton, $g_{\Gamma}$ and $g_{\Gamma,2}$ are g-factors of the $|\pm1_{\Gamma}\rangle$ and $|\pm2_{\Gamma}\rangle$ states, respectively. $g_{e\Gamma}$ is a g-factor of an electron in the $\Gamma$-valley. ${B_{N\parallel}}^{\Gamma}$ and ${B_{N\perp}}^{\Gamma}$ are components of the nuclear field in the $\Gamma$-valley directed parallel and perpendicular to the QDs growth axis.

The matrix form of the indirect exciton block $H_X$:

$$H_X = \frac{\mu_B B}{2}\begin{bmatrix} -g_{X,2} & 0 & 0 & 0 \\ 0 & -g_X & 0 & 0 \\ 0 & 0 & g_X & 0 \\ 0 & 0 & 0 & g_{X,2} \end{bmatrix} + \frac{\mu_B g_{eX}}{2}\begin{bmatrix} {B_{N\parallel}}^{X} & 0 & {B_{N\perp}}^{X} & 0 \\ 0 & {B_{N\parallel}}^{X} & 0 & {B_{N\perp}}^{X} \\ {B_{N\perp}}^{X*} & 0 & -{B_{N\parallel}}^{X} & 0 \\ 0 & {B_{N\perp}}^{X*} & 0 & -{B_{N\parallel}}^{X} \end{bmatrix}, \quad \text{(A2)}$$

here $g_X$ and $g_{X,2}$ are g-factors of the $|\pm1_X\rangle$ and $|\pm2_X\rangle$ states of the indirect exciton, respectively. $g_{eX}$ is the g-factor of an electron in the X-valley. ${B_{N\parallel}}^{X}$ and ${B_{N\perp}}^{X}$ are components of the nuclear field in the X-valley directed parallel and perpendicular to the QDs growth axis.

The matrix form of the $H_{\mathrm{mix}}$ block responsible for the $\Gamma$-X mixing of the electron states:

$$H_{\mathrm{mix}} = \begin{bmatrix} V_{\mathrm{mix}} & 0 & 0 & 0 \\ 0 & V_{\mathrm{mix}} & 0 & 0 \\ 0 & 0 & V_{\mathrm{mix}} & 0 \\ 0 & 0 & 0 & V_{\mathrm{mix}} \end{bmatrix}, \quad \text{(A3)}$$

here $V_{\mathrm{mix}}$ is the strength of the $\Gamma$-X mixing.

The modified block of the indirect exciton $\tilde{H}_X$:

$$\tilde{H}_X = H_X + \frac{{V_{mix}}^2}{2\Delta {E_{\Gamma X}}^2}\begin{bmatrix} -\delta_{0\Gamma} & 0 & 0 & 0 \\ 0 & \delta_{0\Gamma} & \delta_{1\Gamma} & 0 \\ 0 & \delta_{1\Gamma} & \delta_{0\Gamma} & 0 \\ 0 & 0 & 0 & -\delta_{0\Gamma} \end{bmatrix}, \quad \text{(A4)}$$

where $\Delta E_{\Gamma X}$ is the energy splitting between the ground states of the $\Gamma$ and X valleys of the conduction band.

The modified Hamiltonian of the full direct-indirect exciton system:

$$\tilde{H} = \begin{bmatrix} H_{\Gamma} & 0 \\ 0 & \tilde{H}_X \end{bmatrix}. \quad \text{(A5)}$$